\RequirePackage{lineno}
\documentclass[a4paper,aps,12pt,amsmath,amssymb]{revtex4}
\usepackage[dvips]{graphicx}
\usepackage{threeparttable}
\usepackage[yyyymmdd,hhmmss]{datetime}
\usepackage{subfigure}
\usepackage{hyperref}
\usepackage{indentfirst}
\usepackage{color}
\usepackage{rotating}
\begin{document}
\setpagewiselinenumbers
\modulolinenumbers[1]
%\linenumbers

% ---------------------------------------------------------------------------- %
\title{Microstructures and Dynamics of Tetraalkylphosphonium Chloride Ionic Liquids}
\author{Yong-Lei Wang$^{a,b}$$\footnote{Author to whom correspondence should be addressed. Electronic mail: wangyonl@gmail.com}$, Bin Li$^c$, Sten Sarman$^a$, Aatto Laaksonen$^a$}
\affiliation{$^a$Department of Materials and Environmental Chemistry, Arrhenius Laboratory, Stockholm University, SE-106 91 Stockholm, Sweden\\
$^b$Department of Chemistry, Stanford University, Stanford, CA 94305, United States\\
$^c$CAS Key Laboratory for Nanosystem and Hierarchy Fabrication, CAS Center for Excellence in Nanoscience, National Center for Nanoscience and Technology, Chinese Academy of Sciences, Beijing 100190, China}
\date{\today}

% ---------------------------------------------------------------------------- %
\begin{abstract}
Atomistic simulations have been performed to investigate the effect of aliphatic chain length in tetraalkylphosphonium cations on liquid morphologies, microscopic ionic structures and dynamical properties of tetraalkylphosphonium chloride ionic liquids.
The liquid morphologies are characterized by sponge-like interpenetrating polar and apolar networks in ionic liquids consisting of tetraalkylphosphonium cations with short aliphatic chains.
The lengthening aliphatic chains in tetraalkylphosphonium cations leads to polar domains consisting of chloride anions and central polar groups in cations being partially or totally segregated in ionic liquid matrices due to a progressive expansion of apolar domains in between.
Prominent polarity alternation peaks and adjacency correlation peaks are observed at low and high $q$ range in total X-ray scattering structural functions, respectively, and their peak positions gradually shift to lower $q$ values with lengthening aliphatic chains in tetraalkylphosphonium cations.
The charge alternation peaks registered in intermediate $q$ range exhibit complicated tendencies due to the complete cancellations of peaks and anti-peaks in partial structural functions for ionic subcomponents.
The particular microstructures and liquid morphologies in tetraalkylphosphonium chloride ionic liquids intrinsically contribute to distinct dynamics characterized by translational diffusion coefficients, van Hove correlation functions, and non-Gaussian parameters for ionic species in heterogeneous ionic environment.
Most tetraalkylphosphonium cations have higher translational mobilities than their partner anions due to strong coordination of chloride anions with central polar groups in tetraalkylphosphonium cations through strong Coulombic and hydrogen bonding interactions.
The increase of aliphatic chain length in tetraalkylphosphonium cations leads to concomitant shift of van Hove correlation functions and non-Gaussian parameters to larger radial distances and longer timescales, respectively, indicating the enhanced translational dynamical heterogeneities of tetraalkylphosphonium cations and the corresponding chloride anions.
\end{abstract}
\maketitle

% ---------------------------------------------------------------------------- %
\section{Introduction}

\par
Room temperature ionic liquids (ILs) refer to a special category of molten salts entirely composed of bulky asymmetric cations and weakly coordinating anions that exist in liquid state at room temperature.
In recent years, the intense research on ILs has received significant attention in diverse academic and industrial communities due to ILs' multifaceted physicochemical properties, such as non-flammability, negligible volatility, reasonable viscosity-temperature feature, high thermal-oxidative stability, wide electrochemical window, as well as outstanding ability to dissolve polar and apolar compounds~\cite{rogers2003ionic, zhang2006physical, padua2007molecular, greaves2008protic, plechkova2008applications, armand2009ionic, zhou2009ionic, castner2011ionic, hayes2015structure_6357}.
These fascinating characteristics render them reliable alternatives to conventional molecular solvents in industrial applications spanning from synthetic to catalytic chemistry~\cite{rogers2003ionic, greaves2008protic, plechkova2008applications, zhou2009ionic, castner2011ionic, hallett2011room}.
Additionally, their high liquid densities and short Debye screening lengths enable them to effectively screen charged solid surfaces and therefore make them potential replacements of conventional electrolytes in electrochemical energy devices~\cite{armand2009ionic, merlet2013computer_15781, salanne2017ionic, li2016fused}.

\par
A fascinating feature of ILs is that their physicochemical properties, as well as microstructural organization, can be widely tuned in a controllable fashion through combinations of different cation-anion ion pairs in a general way, and by mutating specific atoms in constituent cations or anions~\cite{sanjeevaamurthy2012comparing, kashyap2013does, hayes2015structure_6357, wu2016structure, wu2016structure_2}.
The X-ray~\cite{triolo2007nanoscale, triolo2008morphology, annapureddy2010origin, fujii2011experimental, li2011alkyl, russina2011mesoscopic, santos2011communication, zheng2011effect, shimizu2014structure, liang2015communication, dhungana2016structure, wu2016structure, aoun2011nanoscale, kashyap2012saxs, wu2016structure_2} and neutron scattering~\cite{deetlefs2006liquid, almasy2008structure, triolo2009nanoscale, hardacre2010small, fujii2011experimental, kofu2015quasielastic} experiments, as well as molecular dynamics simulations~\cite{wang2005unique, canongia2006nanostructural, wang2013influence, wang2013multiscale, wang2014heterogeneous, wang2014interfacial, amith2016structures, dhungana2016structure, driver2017correlated, wu2016structure, wu2016structure_2}, have been extensively adopted to elucidate heterogeneous dynamics, microstructural ordering and liquid morphologies of imidazolium and pyrrolidinium based IL systems on nanoscopic level.
These experimental and computational investigations demonstrated an existence of intriguing nanoscopic structural organization in IL matrices.
The liquid structural heterogeneity spans over an order of a few nanometers, and is mainly derived from principle interactions involving different molecular moieties in ILs, that is, Coulombic interactions between charged groups, and dispersive interactions between apolar groups.
Polar groups in imidazolium and pyrrolidinium cations are strongly coordinated with anions, driving to the formation of ionic polar domains from which aliphatic chains are solvophobically excluded, and consequently clustered into apolar domains.
The characteristic size of nanoscopic structural heterogeneities is found to linearly scale with aliphatic chain length in imidazolium and pyrrolidinium cations, which thus provides a sensitive handle for tuning microstructures and physicochemical properties of bulk ILs~\cite{triolo2007nanoscale, triolo2008morphology, annapureddy2010origin, hardacre2010small, fujii2011experimental, kashyap2013structure, shimizu2014structure, araque2015modern}.
The particular liquid morphologies of nanoscopic polar and apolar domains will significantly affect transport behavior in viscosity, diffusion coefficient, and ionic conductivity of ionic species in heterogeneous IL matrices~\cite{canongia2006nanostructural, fayer2014dynamics, wang2014heterogeneous, tamimi2016alkyl, driver2017correlated}.

\par
The heterogeneous microstructures and liquid morphologies are even more distinct in tetraalkylphosphonium based ILs as there are four aliphatic chains in tetraalkylphosphonium cations and each one can be tuned with varied aliphatic substituents~\cite{liu2012microstructures, wang2014atomistic, martak2016new, sun2016effect}, and mutated with different polar and apolar groups~\cite{sanjeevaamurthy2012comparing, kashyap2013does}.
Additionally, tetraalkylphosphonium cations can be associated with various anions and molecular liquids~\cite{gupta2015composition, liang2015communication, wang2015atomistic_1, hettige2016nanoscale, martak2016new, sun2016effect, wang2016atomistic_2, wang2016solvation, dhabal2017structural}, leading to striking ionic structures and distinct liquid morphologies in IL matrices.
Gontrani~\emph{et al.} studied a trihexyltetradecylphosphonium chloride ($[\textrm{P}_{6,6,6,14}]$Cl) IL combining X-ray scattering technique and computer simulation, and alluded the existence of polar segregation at nanoscopic level in IL matrix~\cite{gontrani2009liquid}.
Castner and Margulis groups investigated liquid morphologies and structural ordering properties in ILs consisting of $[\textrm{P}_{6,6,6,14}]$ cations coupled with different anions~\cite{kashyap2012temperature, hettige2016nanoscale}, as well as changes in microstructural landscapes upon heating and pressurizing these ILs~\cite{hettige2014anomalous, sharma2016pressure, dhabal2017structural}.
The liquid organizational morphologies of these $[\textrm{P}_{6,6,6,14}]$ based ILs are dominated by three distinct landscapes at different length scales associated with short range adjacency correlations, positive-negative charge alternations at intermediate range, and long-ranged polarity ordering correlations.
It is the $[\textrm{P}_{6,6,6,14}]$ cation that plays a fundamental role in structuring liquid landscapes of these ILs due to its amphiphilic feature and bulky molecular size.

\par
Since aliphatic chain length is a sensitive handle to microstructural ionic environment, many aspects like how does liquid morphology change in tetraalkylphosphonium based ILs with different aliphatic substituents in cations, and subsequent influence on dynamical properties of ionic species in local ionic environment, are still unclear.
In present work, we performed extensive atomistic simulations to elucidate the effect of linear aliphatic substituents in tetraalkylphosphonium cations on liquid landscapes, microscopic ionic structures and dynamical properties of ionic species in IL matrices.
Six tetraalkylphosphonium cations are considered in present work including triethylbutylphosphonium ($[\textrm{P}_{2,2,2,4}]$), tetrabutylphosphonium ($[\textrm{P}_{4,4,4,4}]$), tributyloctylphosphonium ($[\textrm{P}_{4,4,4,8}]$), tributyltetradecylphosphonium ($[\textrm{P}_{4,4,4,14}]$), tetrahexylphosphonium ($[\textrm{P}_{6,6,6,6}]$), and $[\textrm{P}_{6,6,6,14}]$.
These tetraalkylphosphonium cations are coupled with a monoatomic chloride anion, making it possible to focus on microstructural changes originated from cationic structures without complications introduced by a complex anion.

% ---------------------------------------------------------------------------- %
\section{Ionic Models and Simulation Methodology}

\begin{figure}[!t]
\centering\includegraphics[width=0.5\textwidth]{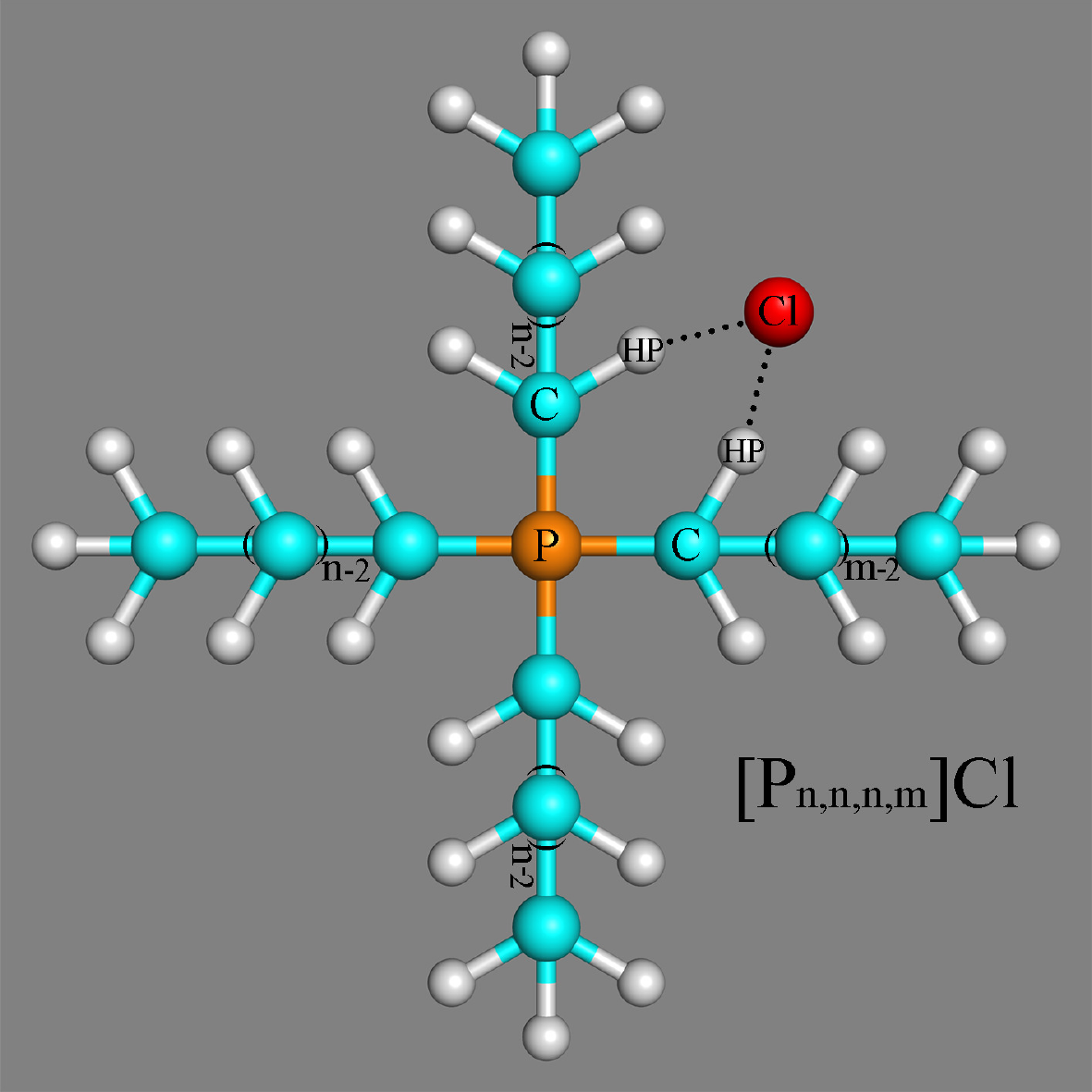}
\caption{Schematic molecular structure of a $[\textrm{P}_{n,n,n,m}]$ cation in coordination with a chloride anion through hydrogen bonding interactions.}\label{fig:structure}
\end{figure}

\par
Molecular structures and representative atom types in tetraalkylphosphonium cation and chloride anion are presented in Fig.~\ref{fig:structure}.
Atomistic force field parameters for six tetraalkylphosphonium chloride ionic liquids are taken from a systematically developed force field in previous works based on AMBER framework~\cite{wang2014atomistic, wang2015multiscale}.
The cross interaction parameters between different atom types are obtained from Lorentz-Berthelot combination rules.
In present atomistic simulations, each simulation system consists of varied number of tetraalkylphosphonium chloride ion pairs with the total atoms of approximately $26000$.
The detailed simulation system compositions are listed in Table~\ref{tbl:table}.
% For $[\textrm{P}_{2,2,2,4}]$Cl, $[\textrm{P}_{4,4,4,4}]$Cl, and $[\textrm{P}_{4,4,4,8}]$Cl ILs, we followed a same procedure to obtain optimized ion pair structures and partial charges on all atoms, and thereafter adopting the same set of interaction parameters as developed in previous works~\ref{wang2014atomistic,wang2015multiscale}.

\begin{table}[t]
\centering\footnotesize
\caption{Simulation system compositions, translational diffusion coefficients (in 10$^{-12}$ m$^2$/s), and peak magnitudes/positions (in ps) of non-Gaussian parameters for cations and anions in six tetraalkylphosphonium chloride ionic liquids at 323 K.}\label{tbl:table}
\begin{tabular}{c@{\quad}c@{\quad}c@{\quad}c@{\quad}c@{\quad}c@{\quad}c@{\quad}c@{\quad}c}
\hline\hline
&\multicolumn{2}{c}{System compositions} & &\multicolumn{2}{c}{Diffusion coefficients} && \multicolumn{2}{c}{Non-Gaussian parameters} \\
\cline{2-3}\cline{5-6}\cline{8-9}
ILs & No. of ion pairs & Total atoms & & Cations & Anions & & Cations & Anions\\
\hline
$[\textrm{P}_{2,2,2,4}]$Cl  & 729 & 26244 && 10.96 & 17.34 && 0.3589/652.1  & 0.2272/342.5 \\
$[\textrm{P}_{4,4,4,4}]$Cl  & 484 & 26136 &&  2.36 & 2.04  && 0.3912/805.8  & 0.2421/562.4 \\
$[\textrm{P}_{4,4,4,8}]$Cl  & 400 & 26400 &&  3.05 & 2.51  && 0.4031/1043.8 & 0.2584/1200.2 \\
$[\textrm{P}_{4,4,4,14}]$Cl & 310 & 26040 &&  3.82 & 3.19  && 0.4293/1585.4 & 0.2847/1828.0 \\
$[\textrm{P}_{6,6,6,6}]$Cl  & 334 & 26052 &&  3.39 & 2.22  && 0.4489/2352.2 & 0.2706/2638.4 \\
$[\textrm{P}_{6,6,6,14}]$Cl & 256 & 26112 &&  3.14 & 1.45  && 0.5135/2535.6 & 0.3554/3698.4 \\
\hline\hline
\end{tabular}
\end{table}

\par
Atomistic molecular dynamics simulations were performed using GROMACS 5.0.4 package~\cite{hess2008gromacs} with cubic periodic boundary conditions.
The equations of motion were integrated using leap-frog integration algorithm with a time step of $1.0$ fs.
A cutoff radius of $1.6$ nm was set for short range van der Waals interactions and real space electrostatic interactions.
The Particle-Mesh Ewald summation method with an interpolation order of $5$ and a Fourier grid spacing of $0.12$ nm was employed to handle long range electrostatic interactions in reciprocal space.
All tetraalkylphosphonium chloride IL simulation systems were first energetically minimized using a steepest descent algorithm, and thereafter annealed gradually from $800$ K to $323$ K within $20$ ns.
The annealed simulation systems were equilibrated in NPT (isothermal-isobaric) ensemble for $60$ ns maintained using Nos\'e-Hoover chain thermostat and Parrinello-Rahman barostat with time coupling constants of $500$ fs and $200$ fs, respectively, to control temperature at $323$ K and pressure at 1 atm.
Atomistic simulations were further performed in NPT ensemble for $80$ ns for all tetraalkylphosphonium chloride ILs, and simulation trajectories were recorded at an interval of $100$ fs for further structural and dynamical analyses.

% ---------------------------------------------------------------------------- %
\section{Results and Discussion}

\subsection{Liquid density}

\begin{figure}[!t]
\centering\includegraphics[width=0.75\textwidth]{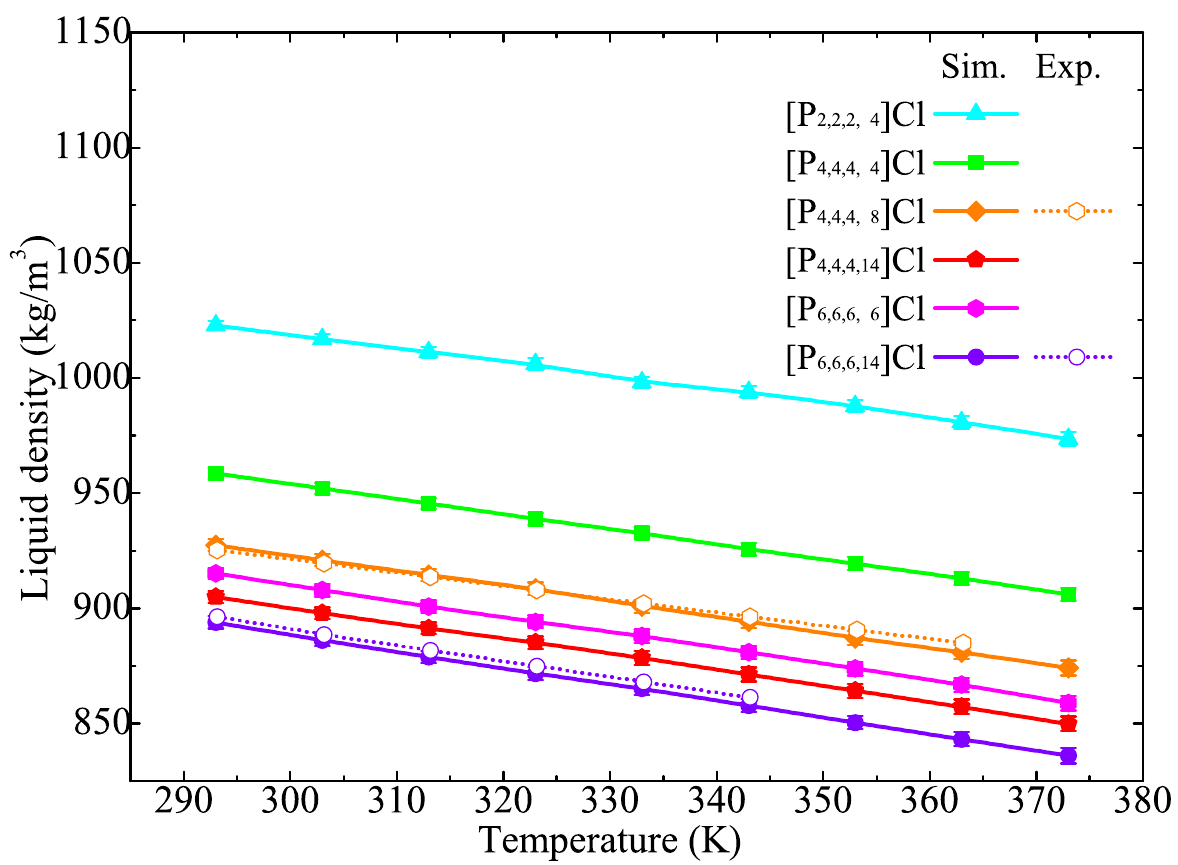}
\caption{Liquid densities of bulk tetraalkylphosphonium chloride ionic liquids at different temperatures obtained from atomistic simulations (solid symbols). The experimental data (open symbols) for $[\textrm{P}_{4,4,4,8}]$Cl and $[\textrm{P}_{6,6,6,14}]$Cl ILs are derived from Refs.~\cite{adamova2012alkyltributylphosphonium} and~\cite{tome2011measurements}, respectively, for comparison.}\label{fig:density}
\end{figure}

\par
Liquid densities of six tetraalkylphosphonium chloride ILs calculated from atomistic simulations at different temperatures ranging from 293 to 373 K are shown in Fig.~\ref{fig:density}.
The experimental density data of $[\textrm{P}_{4,4,4,8}]$Cl and $[\textrm{P}_{6,6,6,14}]$Cl ILs taken from Refs.~\cite{adamova2012alkyltributylphosphonium} and~\cite{tome2011measurements}, respectively, are also provided in Fig.~\ref{fig:density} for comparative propose.
The liquid densities of six ILs decrease with increasing number of carbon atoms in tetraalkylphosphonium cations at specific temperatures.
Additionally, both experimental data and simulation results exhibit linear variations as temperature changes within the range of 293-393 K.
The agreement between experimental data and simulation results is remarkably good over entire temperature range with a maximum deviation of approximately 0.8\% and 1.2\% for $[\textrm{P}_{4,4,4,8}]$Cl and $[\textrm{P}_{6,6,6,14}]$Cl ILs, respectively.
It should be noted that such a small deviation in liquid densities of neat ILs does not show any impact on microstructural and dynamical characterizations as specified in previous publications~\cite{kashyap2013structure, wu2016structure}.

% ---------------------------------------------------------------------------- %
\subsection{Structural function}

\par
To appreciate the overall effect of aliphatic chain length in tetraalkylphosphonium cations on microstructural ordering characteristics in ILs, the total X-ray scattering static structural function, $S(q)$, is calculated using the total sum of atom type based partial components as $S(q) = \sum_{i=1}^n \sum_{j=1}^n S_{ij}(q)$.
The $S_{ij}(q)$ is the partial structural function between atoms types $i$ and $j$ and is given by
\begin{eqnarray}
S_{ij}(q)=\frac{\rho_0 x_i x_j f_i(q) f_j(q) \int_0^{L/2} 4\pi r^2 [g_{ij}(r)-1] \frac{sin(qr)}{qr}W(r)dr}{[\sum_{i=1}^n x_i f_i(q)]^2}\,. \nonumber
\end{eqnarray}
Herein, $g_{ij}(r)$ is the partial radial distribution function between atom types $i$ and $j$, including intra- and intermolecular pairs.
$x_i$ and $x_j$ are the mole fractions of atoms types $i$ and $j$, and $f_i(q)$ and $f_j(q)$ are the corresponding X-ray atomic form factors~\cite{prince2004international}, respectively.
$\rho_0=\frac{N_{atom}}{<L^3>}$ refers to the average atom number density of simulation system and $L$ is the simulation box length.
$W(r)$ is a Lorch window function defined as $W(r) = \frac{sin(2\pi r/L)}{2\pi r/L}$, which is used to minimize the effect of finite truncation of $r$ in the calculation of $g_{ij}(r)$.
Different decomposition scheme of total structural function is advantageous in order to better comprehend the physical origin of striking intermolecular features in $S(q)$ plot.
In practice, the total structural function $S(q)$ can be partitioned into either atomic pair contributions, cationic and anionic subcomponents, or polar and apolar subcomponents, as well as their cross correlations.
The detailed partitioning schemes are systematically described in previous publications and references therein~\cite{hettige2014anomalous, gupta2015composition, dhungana2016structure, sharma2016pressure, wu2016structure_2}.
% For a simulation system of moderately larger size, this window function does not hinder the physical meaning of peaks and antipeaks in $S(q)$ plots.
% The scattering vector $q$ is given by $q=\frac{4\pi sin \theta}{\lambda}$, where $2\theta$ is the angle between the incident X-ray beam and the detector, and $\lambda$ is the X-ray wavelength.

\begin{figure}[!t]
\centering\includegraphics[width=0.8\textwidth]{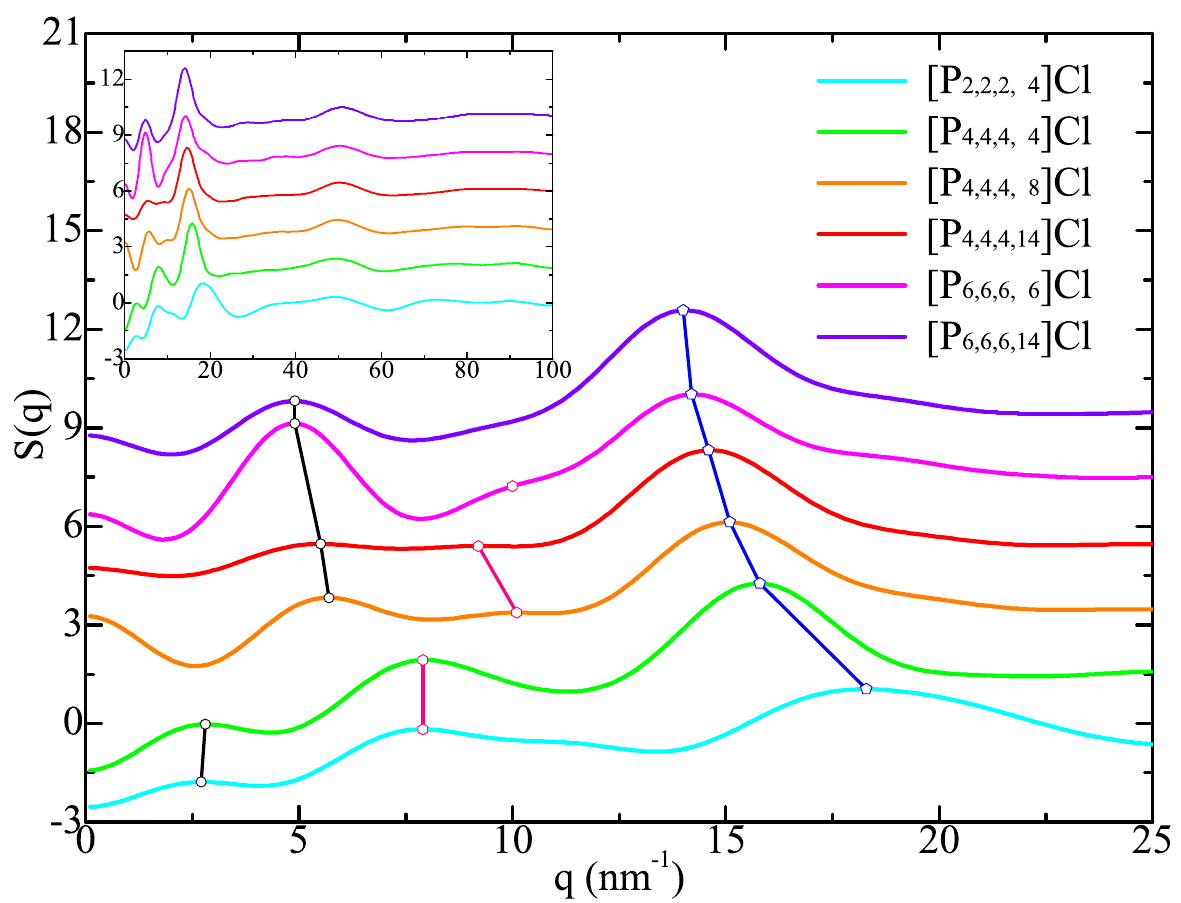}
\caption{Simulated total structural functions $S(q)$ in the range of $q \leq 25.0$ nm$^{-1}$ for six tetraalkylphosphonium chloride ionic liquids at $323$ K. For clarity, these total structural functions are vertically shifted by 2 units based on previous curves for comparative propose. The inset graph depicts the detailed overall structural functions $S(q)$ in the range of $0 < q < 100$ nm$^{-1}$. The peak symbols in low, intermediate and high $q$ values correspond to polar-apolar alternations, positive-negative charge alternations, and close contact adjacency correlations, respectively.}\label{fig:ttsq}
\end{figure}

\par
The total X-ray scattering static structural functions, $S(q)$, for six tetraalkylphosphonium chloride ILs calculated from present atomistic simulations are shown in Fig.~\ref{fig:ttsq}.
Two prominent peaks located at 4.8 nm$^{-1}$ and 13.9 nm$^{-1}$ are shown in total structural function for $[\textrm{P}_{6,6,6,14}]$Cl IL.
These two peak positions are quite consistent with the experimental data positioned at 4.3 nm$^{-1}$ and 13.8 nm$^{-1}$, respectively~\cite{gontrani2009liquid}.
%, which further verify that our computational model is accurate to reproduce the overall experimental structural function.
Such a two-peak-plot is a general feature for $[\textrm{P}_{6,6,6,14}]$ based ILs, and is mainly attributed to close packed ionic structures in local environment~\cite{hettige2016nanoscale, dhabal2017structural}.
The total structural functions for the other five tetraalkylphosphonium chloride ILs generally display three peaks, even the ones located in intermediate $q$ range are not so distinct in some curves.
With some variations across structural function plots for six tetraalkylphosphonium chloride ILs, these peak positions are essential characteristic hallmark of their microstructural landscapes, indicating particular microscopic ionic ordering phenomena at different length scales in IL matrices~\cite{hettige2012anions, kashyap2012saxs, kashyap2013structure, araque2015modern, amith2016structures}.

\begin{figure}[!t]
\centering\includegraphics[width=0.8\textwidth]{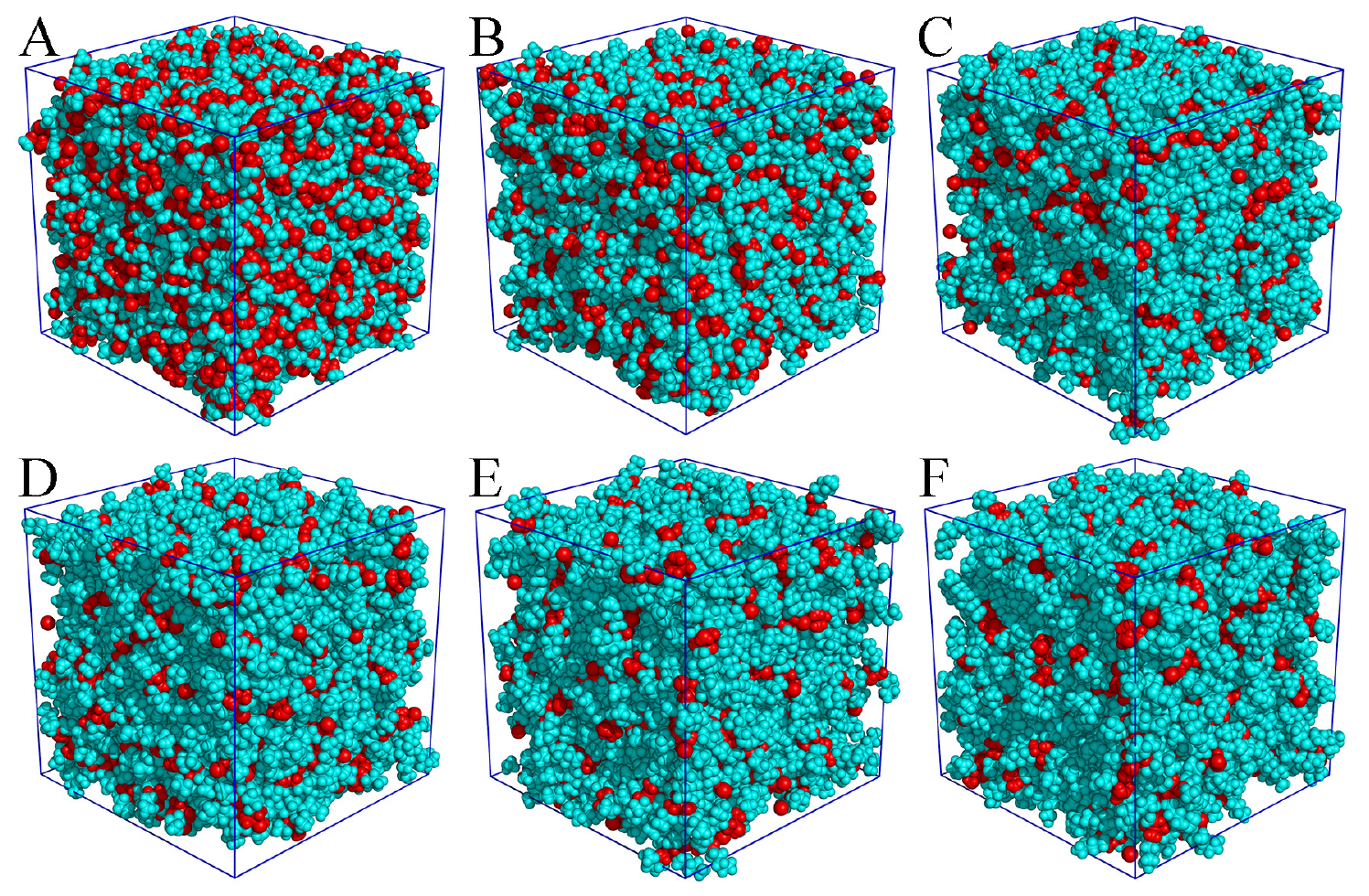}
\caption{Representative liquid morphologies of six tetraalkylphosphonium chloride ionic liquids at 323 K.
(A) $[\textrm{P}_{2,2,2,4}]$Cl; (B) $[\textrm{P}_{4,4,4,4}]$Cl; (C) $[\textrm{P}_{4,4,4,8}]$Cl; (D) $[\textrm{P}_{4,4,4,14}]$Cl; (E) $[\textrm{P}_{6,6,6,6}]$Cl; and (F) $[\textrm{P}_{6,6,6,14}]$Cl.
Polar domains (red) consist of chloride anions and central P(CH$_2$)$_4$ groups in tetraalkylphosphonium cations, and apolar entity (cyan) is composed of the remaining alkyl groups in tetraalkylphosphonium cations, respectively.}\label{fig:morphology}
\end{figure}

\par
The most relevant region in total structural function plots is $q \leq 25.0$ nm$^{-1}$, which is associated with intermolecular correlations between tetraalkylphosphonium cations and the corresponding chloride anions.
Features at $q$ values larger than $25$ nm$^{-1}$ are mostly intramolecular in nature, which are fairly easy to assign and thus are not the subject of this work.
The lowest $q$-peak known as prepeak or First Sharp Diffraction Peak (FSDP), generally occurs around $q=$ 5 nm$^{-1}$, and is an indicative of mesoscopic organization characterized by long range polarity ordering (or polar-apolar density alternation) in IL matrices~\cite{hettige2012anions, kashyap2012saxs, kashyap2013structure, hettige2014anomalous, gupta2015composition, sharma2016pressure}.
The FSDPs in total structural functions for $[\textrm{P}_{2,2,2,4}]$Cl and $[\textrm{P}_{4,4,4,4}]$Cl ILs appear at lower peak values as compared with the counterparts for the other four tetraalkylphosphonium chloride ILs, which are tightly correlated with polar domains encompassing chloride anions and central polar P(CH$_2$)$_4$ groups in tetraalkylphosphonium cations and apolar domains occupied by remaining alkyl groups in aliphatic chains in tetraalkylphosphonium cations.
In $[\textrm{P}_{2,2,2,4}]$Cl and $[\textrm{P}_{4,4,4,4}]$Cl ILs, polar and apolar domain sizes are comparable, and bulk liquid landscapes are characterized by sponge-like or gyroid morphologies with interpenetrating polar and apolar networks, as shown in panels A and B in Fig.~\ref{fig:morphology}.
The increase of aliphatic chain length in tetraalkylphosphonium cations leads to an expansion of apolar network in IL matrices.
In the meantime, the polar network tends to persist but has to accommodate the growing apolar network by loosing part of its connectivity, leading to the segregated distributions of polar domains within apolar framework.
Such a microstructural change in IL matrices contributes to the FSDPs shift from 5.6 nm$^{-1}$ for $[\textrm{P}_{4,4,4,8}]$Cl to 4.8 nm$^{-1}$ for $[\textrm{P}_{6,6,6,14}]$Cl ILs.
Representative liquid morphologies color coded on polar (red) and apolar (cyan) domains for all six tetraalkylphosphonium chloride ILs are shown in Fig.~\ref{fig:morphology}.

\par
The peak at higher $q$ values near $15$ nm$^{-1}$ is associated with short range adjacency correlations originated from nearest neighbouring interactions between ionic species~\cite{gontrani2009liquid, aoun2011nanoscale, santos2011temperature, hettige2012anions, kashyap2013structure, gupta2015composition, hettige2016nanoscale, sharma2016pressure}.
It becomes apparent from our upcoming analysis that this peak mainly originates from apolar adjacency correlations in tetraalkylphosphonium cations.
As expected, significant shifts in peak positions for adjacency correlations toward lower $q$ values (corresponding to larger distances in real space) are observed with lengthening aliphatic chains in tetraalkylphosphonium cations due to the accompanying expansion of apolar network in liquid morphologies as shown in Fig.~\ref{fig:morphology}.

\par
Between the polarity alternation peaks at low $q$ range and the adjacency correlation peaks at high $q$ values, there are peaks at intermediate $q$ range around 10 nm$^{-1}$, which correspond to the positive-negative charge alternations in IL matrices.
This charge ordering behavior is the need to maintain a lattice-like arrangement of cations and anions to minimize Coulombic energy of ionic liquids, and thus is associated with the length scale between ions of the same charge separated by ions of opposite charge~\cite{urahata2004structure, annapureddy2010origin, castner2011ionic, kashyap2012temperature, kashyap2013structure}.
In fact for some ILs, the charge alternation peak is presented only as a weak shoulder, as shown in total structural function plots for $[\textrm{P}_{4,4,4,8}]$Cl, $[\textrm{P}_{4,4,4,14}]$Cl, and $[\textrm{P}_{6,6,6,6}]$Cl ILs in Fig.~\ref{fig:ttsq}.
It is noteworthy that this observation is not because of the missing of charge alternations in these ILs, but instead because of the almost complete cancellations of peaks and anti-peaks that offset this important ordering phenomena at intermediate length scale.
The effect of aliphatic chain length in tetraalkylphosphonium cations on relative positions of charge alternation peaks is complicated, and can be attributed to a gradual transition of polar network from isotropic sponge-like arrangement (panels A and B in Fig.~\ref{fig:morphology}) to partially or totally segregated domains (panels C-F in Fig.~\ref{fig:morphology}) in apolar framework.
This microstructural change in IL matrices has an impact on the number of counterions present in the solvation shells of a given ion, and an even bigger effect on the distributions of (same-charge) ions nearby, and thus contributes to complex tendencies of changes in charge alternation peak positions in intermediate $q$ values.

\par
It is shown in previous works that the total structural functions $S(q)$ can be partitioned in many different ways depending on intended structural properties~\cite{aoun2011nanoscale, hettige2012anions, kashyap2013structure, hettige2014anomalous, amith2016structures, dhungana2016structure, wu2016structure, dhabal2017structural}.
These partitions can be considered as additive projections of the total structural function $S(q)$ onto different subsets of atomic reciprocal space correlations that are chosen to conveniently highlight striking structural properties of ILs.
In following discussion, we mathematically decompose the total structural function into varied subcomponent contributions to address polarity and charge alternations in IL matrices.

\begin{figure}[!t]
\centering\includegraphics[width=0.99\textwidth]{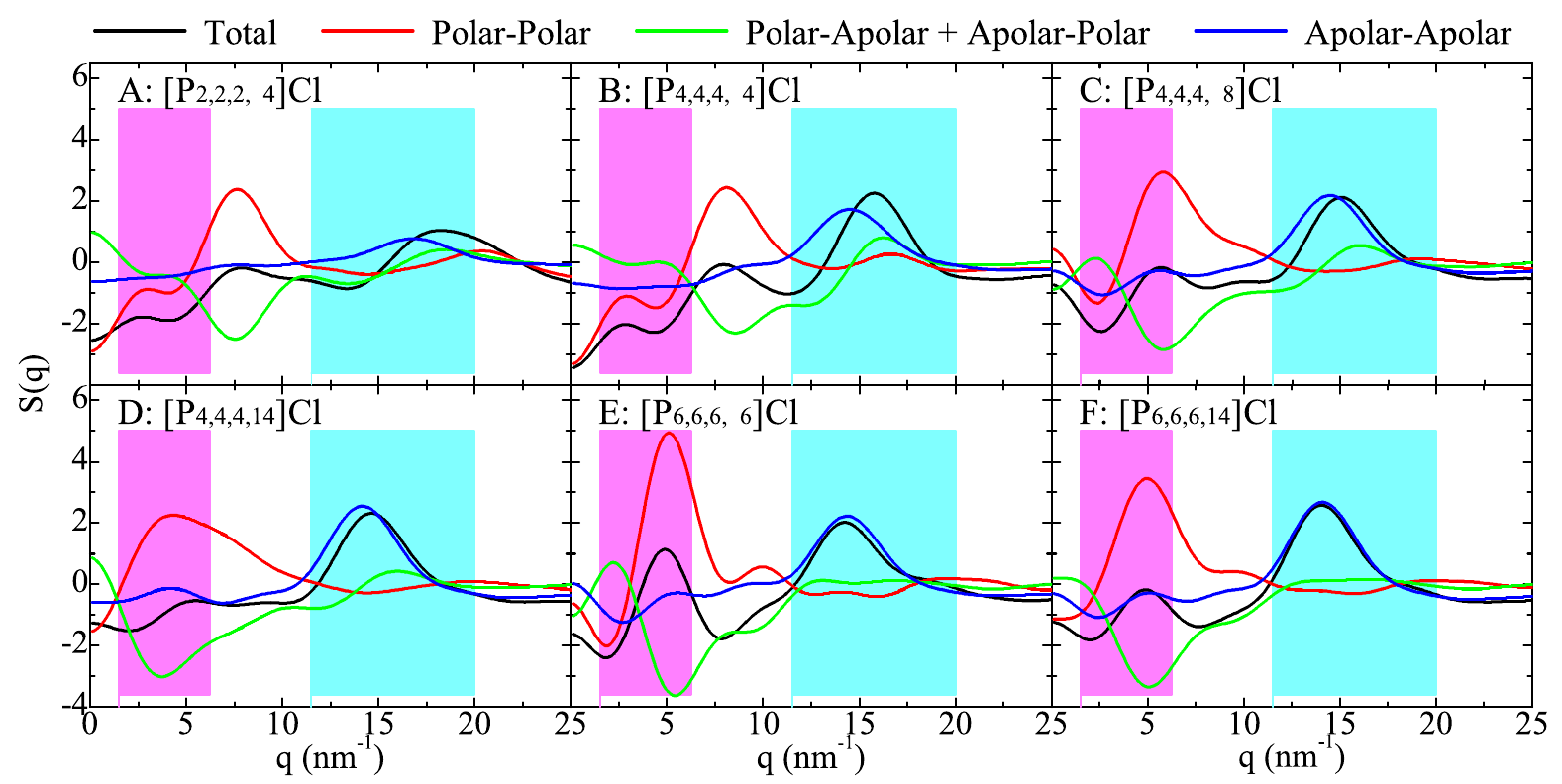}
\caption{Total structural functions $S(q)$ partitioned into polar-polar, polar-apolar/apolar-polar and apolar-apolar subcomponents for six tetraalkylphosphonium chloride ionic liquids at 323 K.}\label{fig:poapsq}
\end{figure}

\par
Fig.~\ref{fig:poapsq} presents the partial structural functions for polar-polar, apolar-apolar, and cross term (polar-apolar/apolar-polar) subcomponents at $q < 25$ nm$^{-1}$, as well as the total structural functions $S(q)$ for six tetraalkylphosphonium chloride ILs.
Herein we focus on the FSDP region colored by magenta, where the polar/apolar partitioning scheme holds special significance.
A prominent prepeak is observed at $q = 2.8$ nm$^{-1}$ in total structural function plots for $[\textrm{P}_{2,2,2,4}]$Cl and $[\textrm{P}_{4,4,4,4}]$Cl ILs, in which the apolar-apolar subcomponent has marginal contribution, and the partial structural functions for polar-polar and polar-apolar/apolar-polar subcomponents are out of sync, respectively.
The onset of FSDPs is shifted to $q = 5$ nm$^{-1}$ in total structural function curves for $[\textrm{P}_{4,4,4,8}]$Cl, $[\textrm{P}_{4,4,4,14}]$Cl, $[\textrm{P}_{6,6,6,6}]$Cl, and $[\textrm{P}_{6,6,6,14}]$Cl ILs.
In these four tetraalkylphosphonium chloride ILs, the polar-polar and apolar-apolar subcomponents contribute to positive intensities, and the cross term exhibits anti-prepeaks at same $q$ values, respectively, indicating a strong polarity ordering in IL matrices.
It is known that the position of FSDPs is associated with typical inverse characteristic distance for polarity alternation ($d=2\pi/q$, where $d$ is the polarity alternation distance in real space)~\cite{aoun2011nanoscale, kashyap2013structure, amith2016structures, wu2016structure, dhabal2017structural}.
The shift of FSDP positions towards lower $q$ values in total structural functions with lengthening aliphatic chains in tetraalkylphosphonium cations indicates that the polarity ordering behavior occurs at longer characteristic distance due to the breakdown of polar network into segregated domains localized within apolar framework.
This explanation is clearly endorsed by typical liquid morphologies of six tetraalkylphosphonium chloride ILs in Fig.~\ref{fig:morphology}.

\par
At higher $q$ values around 15 nm$^{-1}$ colored by cyan, the polar and apolar components have different contributions to adjacency correlation peaks in total structural function plots depending on aliphatic chain length in tetraalkylphosphonium cations.
For tetraalkylphosphonium cations with short aliphatic chains,~\emph{i.e.}, $[\textrm{P}_{2,2,2,4}]$ and $[\textrm{P}_{4,4,4,4}]$, both self and cross subcomponents have obvious contributions to total structural functions $S(q)$.
The lengthening aliphatic chains in tetraalkylphosphonium cations leads to an increased contribution for apolar-apolar subcomponent, and a decreased contribution in polar-polar and polar-apolar/apolar-polar subcomponents, respectively.
Additionally, the difference between the partial structural function for apolar-apolar subcomponent and the total $S(q)$ gradually decreases from $[\textrm{P}_{4,4,4,8}]$Cl to $[\textrm{P}_{6,6,6,14}]$Cl ILs, owing to the strong interference of hydrophobic alkyl groups between segregated polar domains in IL matrices.

\begin{figure}[!t]
\centering\includegraphics[width=0.95\textwidth]{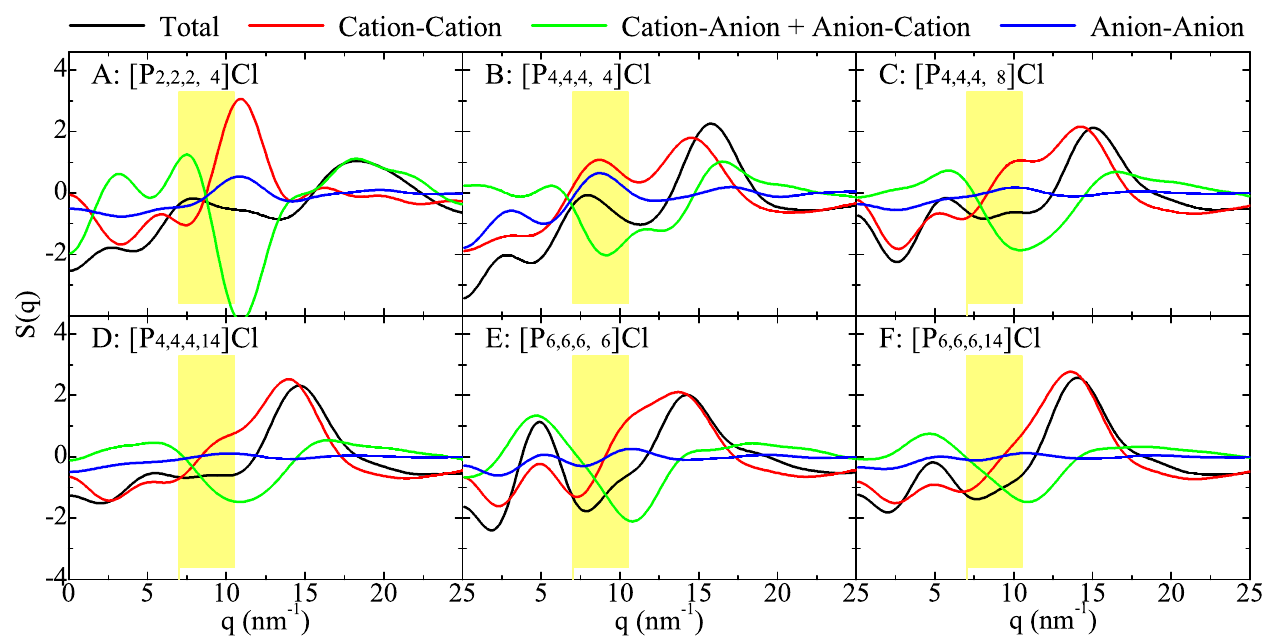}
\caption{Total structural functions $S(q)$ partitioned into cation-cation, cation-anion/anion-cation and anion-anion subcomponents for six tetraalkylphosphonium chloride ionic liquids at 323 K.}\label{fig:caansq}
\end{figure}

\par
Fig.~\ref{fig:caansq} presents ionic partitioning of total structural factors $S(q)$ for six tetraalkylphosphonium chloride ILs at intermediate $q$ range.
The cation-anion/anion-cation subcomponent is always oppositely correlated with cation-cation and anion-anion subcomponents in all six tetraalkylphosphonium chloride ILs, which is a marker of the presence of charge ordering or charge alternation behavior in IL matrices.
The total structural functions for $[\textrm{P}_{2,2,2,4}]$Cl and $[\textrm{P}_{4,4,4,4}]$Cl ILs exhibit clear and prominent peaks at $q = 7.8$ nm$^{-1}$, but have different subcomponent contributions.
In $[\textrm{P}_{2,2,2,4}]$Cl IL, the cation-cation subcomponent exhibits negative contribution, and the cation-anion/anion-cation subcomponent present positive contribution to the total structural function $S(q)$.
However, an opposite effect is observed in partial structural functions for $[\textrm{P}_{4,4,4,4}]$Cl IL, in which the cation-cation and anion-anion subcomponents are characterized by peaks, and the cross term is featured with anti-peak, respectively.
Additionally, it is interesting to observe that the cation-cation and anion-anion subcomponents make strong positive contributions at $q = 10.7$ nm$^{-1}$ to total structural function for $[\textrm{P}_{2,2,2,4}]$Cl IL.
However, these positive contributions are ultimately offseted by anti-correlations between cations and anions at the same $q$ value.
With the increase of aliphatic chain length in tetraalkylphosphonium cations, it is shown that the peaks in partial structural functions for cation-cation and anion-anion subcomponents and the anti-peak in partial structural functions for cation-anion/anion-cation term have different phase and periodicity.
The mathematical additives of these peaks and anti-peaks lead to the descriptive nature of ionic partitioning signature being flatten, and finally diminished in total structural functions for $[\textrm{P}_{4,4,4,8}]$Cl, $[\textrm{P}_{4,4,4,14}]$Cl, $[\textrm{P}_{6,6,6,6}]$Cl, and $[\textrm{P}_{6,6,6,14}]$Cl ILs.
Independent of tetraalkylphosphonium chloride ILs studied in present work, the charge alternation peaks always appear as a result of cancellation of prominent positive contributions from same charge subcomponents and distinct negative-going peaks from cross term contributions.
This observation indicates that at specific locations where one expects to find a same-charge ion there is a systematic absence of ions having opposite charge, which is a generic feature for most ILs~\cite{hettige2012anions, kashyap2012temperature, wu2016structure, dhabal2017structural}.

% ---------------------------------------------------------------------------- %
\subsection{Radial distribution function}

\par
\begin{figure}[!t]
\centering\includegraphics[width=0.95\textwidth]{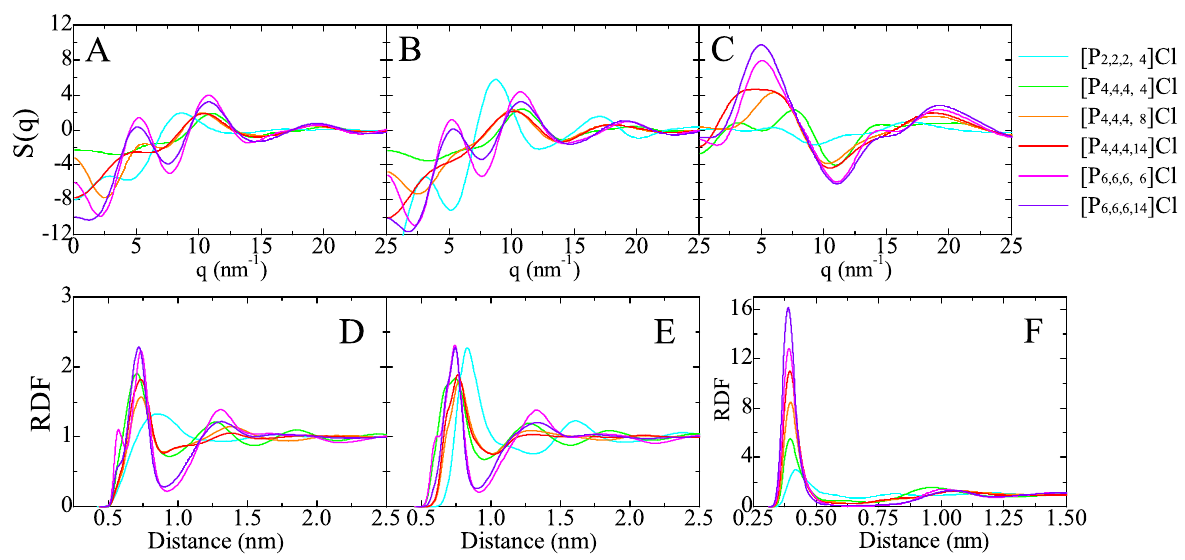}
\caption{Atomic partial structural functions and radial distribution functions of P-P (A and D), Cl-Cl (B and E) and P-Cl (C and F) pairs in six tetraalkylphosphonium chloride ionic liquids at 323 K.}\label{fig:pcl_sqrdf}
\end{figure}

To look further into the real space correlations between tetraalkylphosphonium cations and chloride anions, we calculate the site-site radial distribution functions (RDFs) between chloride anions and the central phosphorous (P) atoms in tetraalkylphosphonium cations.
The P-P, P-Cl and Cl-Cl pair RDFs, and the corresponding partial structural functions, are shown in Fig.~\ref{fig:pcl_sqrdf}.
For $[\textrm{P}_{2,2,2,4}]$Cl IL, both P-P and Cl-Cl RDFs exhibit different forms to their counterparts for the other five ILs, due to the intrinsic liquid morphologies and microscopic ionic structures in $[\textrm{P}_{2,2,2,4}]$Cl IL matrix.
The lengthening aliphatic chains in tetraalkylphosphonium cations leads to a shift of the principal peaks in these two PDF plots to short radial distances.
These microstructural changes between chloride anions and central phosphorous atoms in tetraalkylphosphonium cations in polar domains are qualitatively manifested in corresponding partial structural functions.

\par
The P-Cl RDFs shown in Fig.~\ref{fig:pcl_sqrdf}F present negligible changes in the first peak positions around 0.40 nm, indicating that the spherical chloride anions display qualitatively similar distributions around tetraalkylphosphonium cations in all six IL simulation systems.
However, an enhanced short range correlation is observed in P-Cl RDF plots with lengthening aliphatic chains in tetraalkylphosphonium cations, which is intrinsically related with liquid morphologies and microstructural changes in IL matrices.
As discussed in previous subsections, the liquid morphologies are gradually transitioned from sponge-like interpenetrating polar and apolar networks for $[\textrm{P}_{2,2,2,4}]$Cl IL to totally segregated polar domains within apolar framework for $[\textrm{P}_{6,6,6,14}]$Cl IL.
During such a liquid morphology transition, P and Cl atoms are decisively coordinated in the first solvation shell in polar domains through strong Coulumbic interactions.
However, the intermolecular interactions between P and Cl atoms beyond the first solvation shell are partially or totally screened due to spatial segregation of polar domains within apolar framework.
These two factors eventually promote the enhanced solvation of chloride anions around central polar groups in tetraalkylphosphonium cations.
Such a particular ionic environment between close contacted P and Cl atoms are qualitatively reflected in distinctive peaks around 4.9 nm$^{-1}$ and 19.2 nm$^{-1}$, and anti-peaks at 11.1 nm$^{-1}$, respectively, in partial structural functions for P-Cl pair in Fig.~\ref{fig:pcl_sqrdf}C.

\par
Pair correlation functions between terminal carbon atoms of aliphatic chains in tetraalkylphosphonium cations exhibit imperceptible changes in peak heights and peak positions (data not shown).
This observation indicates that molecular size of tetraalkylphosphonium cations has little influence on hydrophobic interactions and local structures of terminal carbon atoms of aliphatic chains in tetraalkylphosphonium cations.
Terminal carbon atoms are isotropically distributed and exhibit mean coordination with neighboring hydrophobic alkyl groups within apolar domains.

\begin{figure}[!t]
\centering\includegraphics[width=0.6\textwidth]{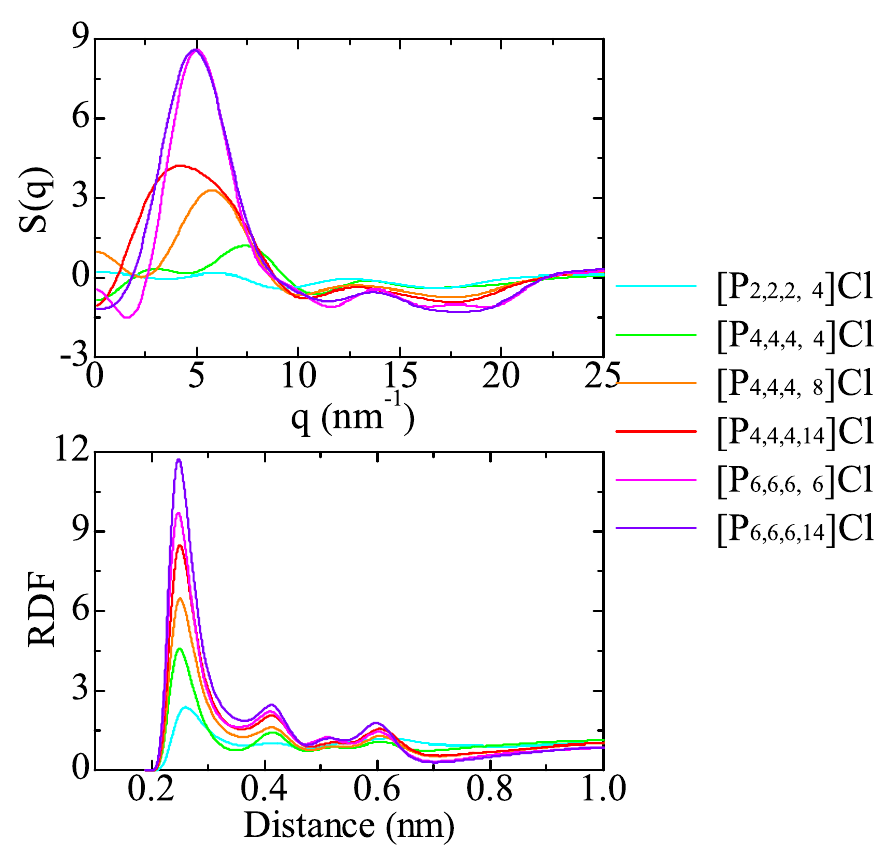}
\caption{Atomic partial structural functions and radial distribution functions of HP-Cl pair in six tetraalkylphosphonium chloride ionic liquids at 323 K.}\label{fig:hpcl_sqrdf}
\end{figure}

\par
Hydrogen bonding is an important feature in ILs as it incorporates many distinct characteristics in their microstructural landscapes and dynamics.
For tetraalkylphosphonium based ILs, there is only one type of hydrogen bond linkage between tetraalkylphosphonium cations and the coupled chloride anions, as verified utilizing geometric criterion based on distance and angular constraints in previous work~\cite{liu2012microstructures, wang2014atomistic, wang2015multiscale}.
In present work, the intermolecular RDFs and partial structural functions between HP (as labelled in Fig.~\ref{fig:structure}) and Cl atoms in six tetraalkylphosphonium chloride ILs are shown in Fig.~\ref{fig:hpcl_sqrdf}.
Similar microstructural feature is observed in HP-Cl RDF plots as that for P-Cl pair RDFs.
This is expected as P and HP atoms are indirectly correlated through the P-C-HP angle in tetraalkylphosphonium cations.
The chloride anions are always coordinated with central polar P(CH$_2$)$_4$ groups due to attractive Coulombic forces in P-Cl pair and specific hydrogen bonding interactions between HP and Cl atoms.
The synergistic effect of these intermolecular interactions promotes a constrained spatial distribution of chloride anions in tetrahedral regions formed by four aliphatic chains in tetraalkylphosphonium cations.

% ---------------------------------------------------------------------------- %
\subsection{Translational dynamics}

\begin{figure}[!t]
\centering\includegraphics[width=0.95\textwidth]{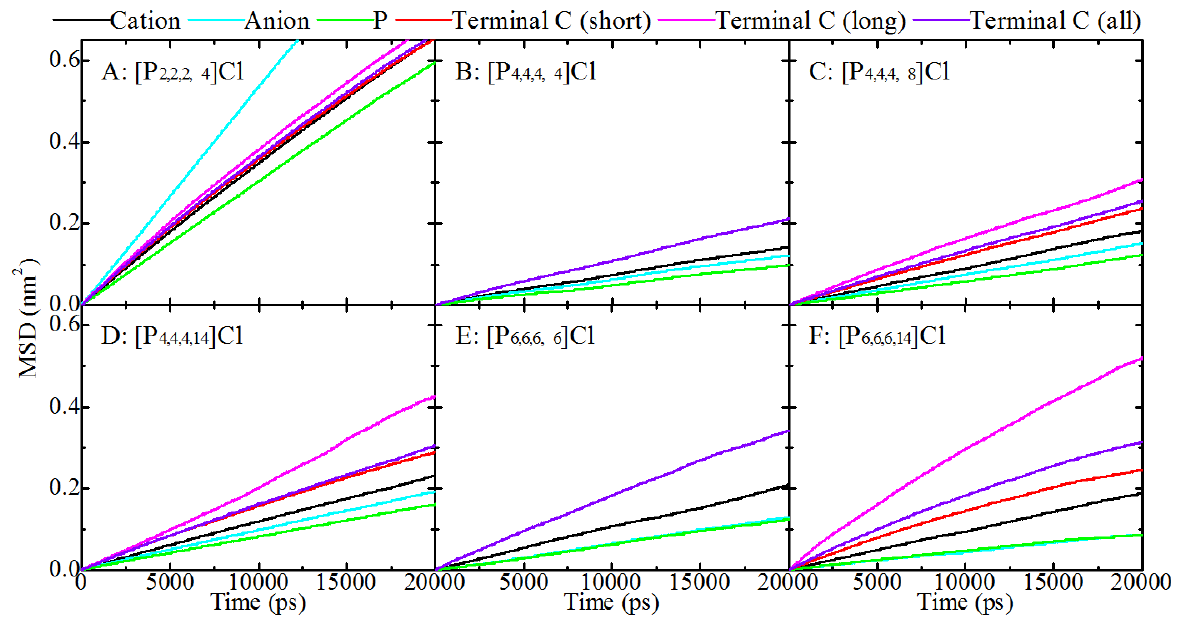}
\caption{Mean square displacements of chloride anions and representative atoms in tetraalkylphosphonium cations including central phosphorous atoms, terminal carbon atoms in short and long aliphatic chains, and centers-of-mass of cations in six tetraalkylphosphonium chloride ionic liquids at 323 K.}\label{fig:msd}
\end{figure}

\par
Quantitative characterization of translational dynamics of tetraalkylphosphonium cations and chloride anions is quantified in terms of mean square displacement (MSD), which is expressed as:
\begin{eqnarray}
\textrm{MSD}(t)=\frac{1}{N}\sum_{i=1}^N \Big\langle [r_i(t)-r_i(0)]^2 \Big\rangle\,,
\end{eqnarray}
where $N$ is the number of ion type $i$ in simulation box, and $r_i(t)$ is the center-of-mass (COM) coordinate of ion $i$ at a given time $t$.
The angular bracket $\langle\rangle$ represents an ensemble average over multiple time origins and all same type ions in simulation box to improve statistical precision.
The time variations of MSDs in diffusive regime obtained from NPT simulations at 323 K are shown in Fig.~\ref{fig:msd} for chloride anions and COMs of tetraalkylphosphonium cations, as well as that for central P atoms and terminal carbon atoms of aliphatic chains in tetraalkylphosphonium cations.
Self diffusion coefficients of tetraalkylphosphonium cations and chloride anions are determined from MSD plots using Einstein relation, $D=\frac{1}{6}\displaystyle{\lim_{t\rightarrow\infty}}\frac{d}{dt}\Big(\textrm{MSD}(t)\Big)$, and are listed in Table~\ref{tbl:table}.
The magnitude of diffusion coefficients for tetraalkylphosphonium cations and chloride anions at $323$ K is $10^{-12}$ m$^2$/s, which is a common order for translational diffusion of ionic species at similar temperature range~\cite{tokuda2005physicochemical,liu2011molecular,wang2015multiscale}.
The terminal carbon atoms in long aliphatic chains in tetraalkylphosphonium cations exhibit faster diffusion than that for their counterpart in short aliphatic chains, as shown in panels A, C, D, and F in Fig.~\ref{fig:msd}.
The four aliphatic chains in $[\textrm{P}_{4,4,4,4}]$ and $[\textrm{P}_{6,6,6,6}]$ cations are characterized with same length and the four terminal carbon atoms in aliphatic chains have almost the same MSD curves (data not shown), and thus their average MSD data are present in panels B and E in Fig.~\ref{fig:msd}.
For tetraalkylphosphonium chloride ILs with an exception of $[\textrm{P}_{2,2,2,4}]$Cl, the translational diffusions of specific groups follow an order of terminal C$_{\textrm{long~chains}}$ $>$ terminal C$_{total}$ $>$ terminal C$_{\textrm{short~chains}}$ $>$ tetraalkylphosphonium cations $>$ P atoms in tetraalkylphosphonium cations.
It is the fast diffusion of terminal carbon atoms in aliphatic chains and the relative slow mobility of central P atoms that lead to the medium self diffusivity of the whole tetraalkylphosphonium cations.

\par
It is shown in Fig.~\ref{fig:msd} that the translational diffusion of tetraalkylphosphonium cations is concerted with that of the corresponding chloride anions, as indicated by comparable MSD plots for $[\textrm{P}_{4,4,4,4}]$Cl, $[\textrm{P}_{4,4,4,8}]$Cl, and $[\textrm{P}_{4,4,4,14}]$Cl ILs and the overlap of MSD curves for $[\textrm{P}_{6,6,6,6}]$Cl and $[\textrm{P}_{6,6,6,14}]$Cl ILs, respectively.
This is striking as tetraalkylphosphonium cations are quite bulky and voluminous than simple chloride anions.
Compared with imidazolium cations which have preferential translational diffusion along the imidazolium plane, the four aliphatic chains in tetraalkylphosphonium cations are characterized with distinct group sizes.
In $[\textrm{P}_{4,4,4,14}]$ and $[\textrm{P}_{6,6,6,14}]$ cations, the tetradecyl chain is longer than the other three linear substituents, but this structural discrepancy does not lead to preferential translational motion of the whole cation along the tetradecyl chain direction.
This observation can be intrinsically attributed to constrained microstructural coordination of chloride anions with central polar P(CH$_2$)$_4$ groups in tetraalkylphosphonium cations in heterogeneous IL matrices.
The spatial constraint distributions of tetraalkylphosphonium cations and chloride anions in IL matrices have threefold structural features.
First is the microstructural constraint induced by strong electrostatic and preferential hydrogen bonding interactions.
Tetraalkylphosphonium cations, especially the central polar P(CH$_2$)$_4$ groups, are constrained by surround chloride anions through attractive electrostatic interactions between P and Cl atoms, and through multiple directional hydrogen bonding interactions between HP and Cl atoms.
Next is the steric constraint originated from aliphatic chains in tetraalkylphosphonium cations.
The chloride anions, on the one hand, are constrained by P and HP atoms due to preferential intermolecular interactions, and on the other hand, their rear parts are wrapped by methyl groups in aliphatic chains, leading to their steric constraint in small cavities in tetrahedral regions of tetraalkylphosphonium cations.
The last one is the domain constraint due to relative distribution of polar and apolar domains in IL matrices.
The microstructural landscapes of polar domains are characterised by sponge-like interpenetrating networks in ILs consisting of small tetraalkylphosphonium cations, like $[\textrm{P}_{2,2,2,4}]$, and by partially connected morphologies in IL matrices consisting of intermediate tetraalkylphosphonium cations, like $[\textrm{P}_{4,4,4,4}]$, $[\textrm{P}_{4,4,4,8}]$ and $[\textrm{P}_{4,4,4,14}]$, and eventually transitioned to segregated distributions in apolar framework consisting of methyl groups in $[\textrm{P}_{6,6,6,6}]$ and $[\textrm{P}_{6,6,6,6}]$ cations, respectively.
Such a trifold effect leads to distinct translational diffusivity of representative atoms in tetraalkylphosphonium cations and chloride anions in heterogeneous ionic environment.
In $[\textrm{P}_{2,2,2,4}]$Cl IL, the microstructural constraint works but not for the other two constraints, and thus the self diffusivity of chloride anions are higher than any part of, and even the whole $[\textrm{P}_{2,2,2,4}]$ cations.
In $[\textrm{P}_{4,4,4,4}]$Cl, $[\textrm{P}_{4,4,4,8}]$Cl and $[\textrm{P}_{4,4,4,14}]$Cl ILs, the cooperative effect of microstructural and steric constricts leads to the comparable translational mobilities of P and Cl atoms.
But the butyl chains in these three tetraalkylphosphonium cations are not long enough to sterically constrain chloride anions within their local tetrahedral regions, resulting in the self diffusivity of chloride anions being always 10\% higher than that for P atoms in corresponding IL matrices.
In $[\textrm{P}_{6,6,6,6}]$Cl and $[\textrm{P}_{6,6,6,6}]$Cl ILs, the synergistic effect of trifold constraints contributes to the overlap of MSD plots for P and Cl atoms, indicating that the hexane chains can spatially and dynamically constrain chloride anions within their local cavities in an effective procedure.

\par
It is interesting to notice that the translational mobility of terminal carbon atoms, either in long or in short aliphatic chains, and their total diffusion increase with tetraalkylphosphonium cationic molecular sizes, as clearly shown in panels B-F in Fig.~\ref{fig:msd}.
This observation indicates that apolar network moves faster with lengthening aliphatic chains in tetraalkylphosphonium cations.
However, the diffusion of polar domains, mainly the mobilities of P and Cl atoms, presents complicated tendencies.
The translational diffusion coefficients for P and Cl atoms first increase with lengthening aliphatic chains from $[\textrm{P}_{4,4,4,4}]$Cl, $[\textrm{P}_{4,4,4,8}]$Cl to $[\textrm{P}_{4,4,4,14}]$Cl ILs, but then decrease for $[\textrm{P}_{6,6,6,6}]$Cl and $[\textrm{P}_{6,6,6,14}]$Cl ILs.
The striking dynamics of polar and apolar domains are essentially related with the gradual addition of methyl groups in tetraalkylphosphonium cations.
On the one hand, the addition of a methyl (-CH$_2$-) unit to tetraalkylphosphonium cations weakens the electrostatic attractions between tetraalkylphosphonium cations and chloride anions, leading to the enhanced mobilities of P and Cl atoms, as well as that for terminal carbon atoms in aliphatic chains and the whole tetraalkylphosphonium cations as shown in panels B-D in Fig.~\ref{fig:msd}.
However, on the other hand, the addition of methyl units to tetraalkylphosphonium cations enhances the van der Waals interactions by means of aliphatic chain-ion inductive forces and hydrophobic interactions between methyl groups.
This will increase the self diffusivity of terminal carbon atoms in aliphatic chains, which, however, will in turn compress the translational dynamics of polar domains due to their segregated distributions in apolar framework, as shown in panels E-F in Fig.~\ref{fig:msd}.
In present work, it is the delicate interplay of Coulombic interactions between chloride anions and central polar groups in tetraalkylphosphonium cations and dispersion interactions among remaining alkyl units in tetraalkylphosphonium cations that results in the striking translational dynamics of ionic groups in IL matrices.

% ---------------------------------------------------------------------------- %
\subsection{Dynamical heterogeneity}

\begin{figure}[!t]
\centering\includegraphics[width=0.95\textwidth]{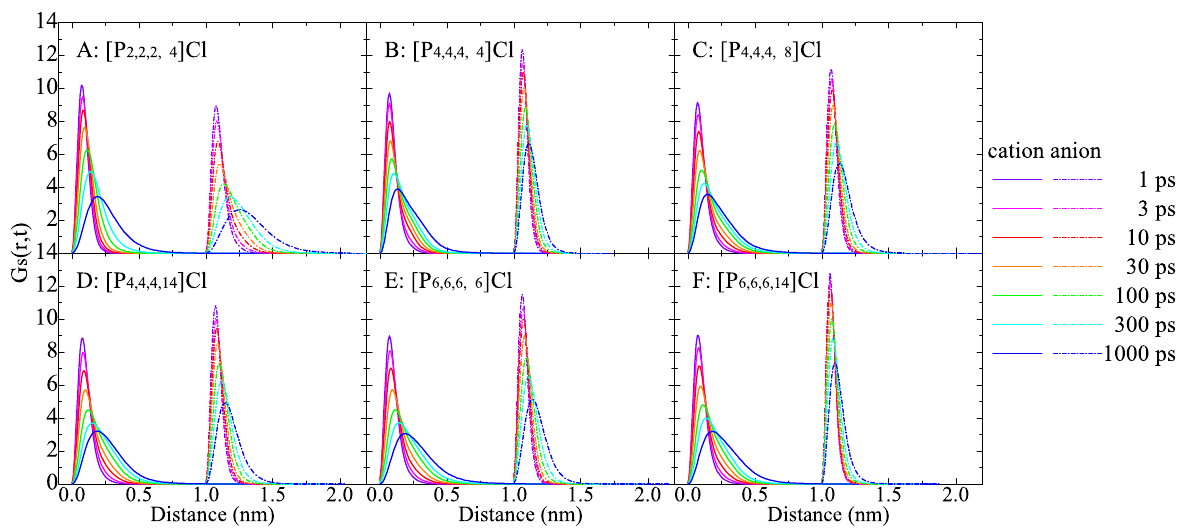}
\caption{Self part of van Hove correlation functions $G_s(r,t)$ for tetraalkylphosphonium cations and chloride anions at seven different times at 323 K. The functions for chloride anions are horizontally shifted by 1 nm for clarity and comparative propose.}\label{fig:gs}
\end{figure}

\par
Atomistic simulation results in previous subsections indicate that translational dynamics of tetraalkylphosphonium cations and chloride anions are heterogeneous depending on aliphatic chain length in tetraalkylphosphonium cations.
Since MSD plots give typical average distance that a tagged ionic group moves within a time $t$, it is of interest to investigate the distribution of its displacement, which can be quantified by computing time dependent self part of van Hove correlation function $G_s(r,t)$ as~\cite{hu2006heterogeneity, singh2011heterogeneity, liu2011molecular, wang2014heterogeneous}:
\begin{eqnarray}
G_s(r,t)=\frac{1}{N}\sum_{i=1}^N\langle\delta[r+r_i(t)-r_i(0)]\rangle\,.\nonumber
\end{eqnarray}
Fig.~\ref{fig:gs} presents the self part of van Hove correlation functions for tetraalkylphosphonium cations and corresponding chloride anions at seven different times at $323$ K.
All these van Hove correlation functions have a single peak and its peak position moves quickly to longer distance as time increases.
The van Hove correlation functions at very short times are characterized by Gaussian type distributions indicating a ballistic motion of ionic species in local environment, but significant deviations from standard Gaussian form are obvious at intermediate times.
This observation indicates that most ionic groups are temporarily trapped within a cage formed by surrounding counterions, and exhibit slower translational diffusion than that expected from the Fick\rq s law.
The deviation of van Hove correlations functions from Gaussian type distribution,~\emph{i.e.}, at $1000$ ps, suggests that the trapped ionic groups have not yet left the initial cage formed by their surrounding counterions at time $t=0$ ps.
In each tetraalkylphosphonium chloride IL with an exception of $[\textrm{P}_{2,2,2,4}]$Cl IL, the van Hove correlation functions for tetraalkylphosphonium cations are characterized with broad distributions at larger radial distances than the corresponding chloride anions at the same time interval, indicating that cations have larger translational mobilities than chloride anions as clearly verified in MSD plots in Fig.~\ref{fig:msd}.

\par
Only after sufficient long time can these trapped ionic groups escape their cages, and the corresponding simulation systems reach diffusive regime, in which the self diffusivities of ionic groups are characterized by MSD $\propto\Delta t$ as shown in Fig.~\ref{fig:msd}.
Additionally, these van Hove correlation functions exhibit long tails due to some ions diffusing faster than mainstream diffusion in IL matrices, but no activated~\lq\lq hopping\rq\rq~process occurring at 323 K in present atomistic simulations.
This observation indicates that the translational diffusion of tetraalkylphosphonium cations and chloride anions at 323 K is a continuous process controlled by the time it takes to leave a trapped cage of its surrounding counterions without entering another one due to the missing of secondary peaks in van Hove correlation functions at large distance and long time scales~\cite{scheidler2004relaxation, hu2006heterogeneity, singh2011heterogeneity, liu2011molecular, wang2014heterogeneous}.

\begin{figure}[!t]
\centering\includegraphics[width=0.75\textwidth]{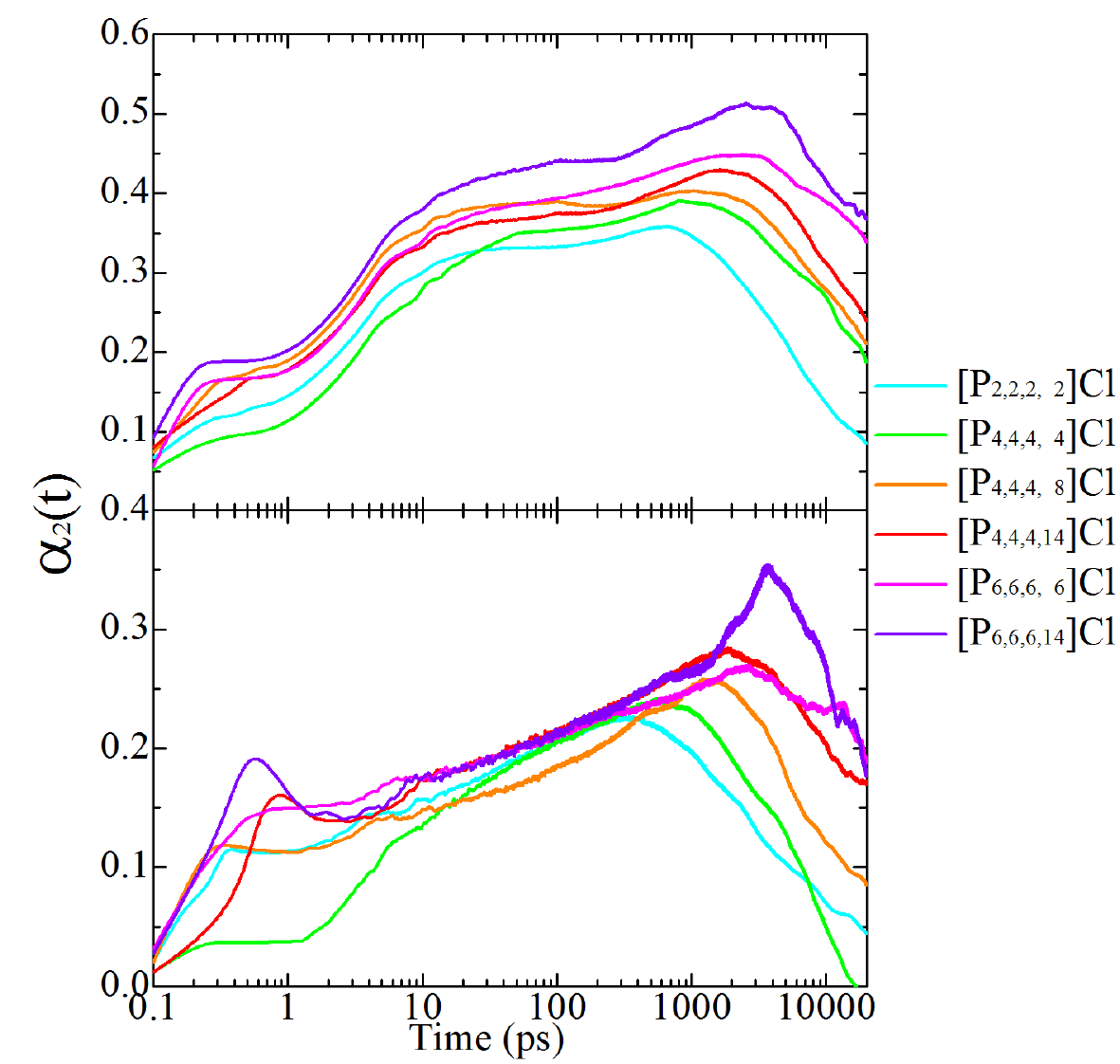}
\caption{Non-Gaussian parameters $\alpha_2(t)$ for tetraalkylphosphonium cations (upper panel) and chloride anions (lower panel) at 323 K.}\label{fig:alpha}
\end{figure}

\par
The deviation of translational mobilities of tetraalkylphosphonium cations and chloride anions from Gaussian behavior is associated with dynamic heterogeneity and can be quantified by non-Gaussian parameter defined as~\cite{del2004structure,hu2006heterogeneity,liu2011molecular}:
\begin{eqnarray}
\alpha_2(t) = \frac{3\langle r^4(t)\rangle}{5\langle r^2(t)\rangle^2} - 1\,,\nonumber
\end{eqnarray}
where $r(t)$ is the displacement of an ionic group at time $t$ with respect to its position at time $t=0$.
Fig.~\ref{fig:alpha} presents the non-Gaussian parameters $\alpha_2(t)$ for tetraalkylphosphonium cations and the corresponding chloride anions at 323 K.
The non-Gaussian parameter depicts a nonmonotonic time dependence with peak positions at different time scale.
At short times, non-Gaussian parameter is zero due to a free particle behavior of ionic groups in their initial temporal regime.
The non-Gaussian parameter increases with time and reaches a maximum at intermediate time.
This maximum corresponds to a timescale at which the trapped ionic groups with fastest translational mobilities brake the cage of surrounding counterions and enter diffusive regimes.
The peak intensities and peak positions in non-Gaussian parameter curves for tetraalkylphosphonium cations and chloride anions, as listed in Table~\ref{tbl:table}, become stronger and shift to larger timescale with lengthening aliphatic chains in tetraalkylphosphonium cations.
This observation is attributed to strong microstructural and dynamical heterogeneities of constrained tetraalkylphosphonium cations and chloride anions in heterogeneous IL matrices.
It is noteworthy that the peak intensities and peak positions in non-Gaussian parameter plots for tetraalkylphosphonium cations are higher and larger than that for their partner anions, indicating that tetraalkylphosphonium cations have a higher degree of dynamical heterogeneity, which mainly comes from the almost uniform distribution of chloride anions around tetraalkylphosphonium cations.

\par
At times longer than peak timescales, non-Gaussian parameters will decay due to diffusive motion of ionic species in IL matrices.
It is expected that the non-Gaussian parameter $\alpha_2(t)$ should be recovered to zero for dynamically homogeneous system.
However, such a process is very slow for almost all IL systems due to the overall heterogeneous structures and dynamics of ionic species.
The decay of chloride anions is much faster than that for corresponding tetraalkylphosphonium cations after peak timescales, as expected.
We can infer from Fig.~\ref{fig:alpha} that the non-Gaussian parameters for both tetraalkylphosphonium cations and chloride anions in all six simulation systems are finite owing to the heterogeneous microstructures and dynamics of ionic species in local ionic environment.
Actually, the deviation of non-Gaussian parameter $\alpha_2(t)$ from zero is an indicative of extent and timescale of relaxation process of ionic species with different reorientational rates in heterogeneous ionic matrices~\cite{del2004structure, liu2011molecular}.

% ---------------------------------------------------------------------------- %
\section{Concluding Remarks}

\par
Atomistic simulations have been performed to investigate the effect of aliphatic chain length in tetraalkylphosphonium cations on liquid morphologies, microscopic ionic structures and dynamical properties of ionic species in tetraalkylphosphonium chloride ILs.
Detailed analyses of total and partial structural functions and radial distribution functions between specific atoms in tetraalkylphosphonium cations and chloride anions indicate that bulk ILs are characterized by distinct microscopic ionic structures and heterogeneous liquid morphologies depending on aliphatic chain length in tetraalkylphosphonium cations.
For ILs consisting of small tetraalkylphosphonium cations, like in $[\textrm{P}_{2,2,2,4}]$Cl IL, microstructural liquid morphologies are characterized by bicontinuous sponge-like interpenetrating polar and apolar networks.
The lengthening aliphatic chains in tetraalkylphosphonium cations leads to the polar network being partially broken or totally segregated within apolar framework so as to accommodate progressive expansion of apolar domains.

\par
The liquid landscape variations and heterogeneous microstructural changes in six tetraalkylphosphonium chloride ILs are qualitatively verified by prominent polarity alternation peaks and adjacency correlation peaks observed at low and high $q$ range in total structural functions, respectively, and their peak positions gradually shift to lower $q$ values with lengthening aliphatic chains in tetraalkylphosphonium cations.
The effect of aliphatic chain length in tetraalkylphosphonium cations on relative positions of charge alternation peaks registered in intermediate $q$ range is complicated because of the perfect cancellations of positive contributions of same charge ions and negative contributions of ions of opposing charge.

\par
The particular liquid morphologies and heterogeneous microscopic ionic structures in tetraalkylphosphonium chloride ILs are intrinsically manifested in dynamical properties characterized by mean square displacements, translational mobilities, time dependent self part of van Hove correlation functions, and non-Gaussian parameters of tetraalkylphosphonium cations and chloride anions in bulk liquid systems.
The terminal carbon atoms in aliphatic chains exhibit overall higher diffusivity than central P atoms in tetraalkylphosphonium cations, and their cooperative effect contributes to the medium diffusion coefficients of the whole tetraalkylphosphonium cations.
The P and Cl atoms exhibit comparable translational diffusivities due to their strong Coulombic coordination feature in polar domains highlighting the existence of strongly correlated ionic structures in IL matrices.
The lengthening aliphatic chains in tetraalkylphosphonium cations leads to concomitant shift of van Hove correlation functions and non-Gaussian parameters to larger radial distances and longer timescales, respectively, indicating the enhanced translational dynamical heterogeneities of tetraalkylphosphonium cations, as well as that for corresponding chloride anions in constrained local environment.

% ---------------------------------------------------------------------------- %
\section*{Acknowledgment}
Y.-L. Wang gratefully acknowledges financial support from Knut and Alice Wallenberg Foundation (KAW 2015.0417). A. Laaksonen acknowledges Swedish Science Council for financial support. All atomistic simulations were performed using computational resources provided by Swedish National Infrastructure for Computing (SNIC) at PDC, HPC2N and NSC.

% ---------------------------------------------------------------------------- %
\bibliography{wang_pxxxxcl_ms}

\end{document}